\documentclass[%
preprint,
]{revtex4}

\usepackage{graphicx}% Include figure files
\usepackage{dcolumn}% Align table columns on decimal point
\usepackage{bm}% bold math

\usepackage{mathrsfs}

\begin{document}
\title{Semileptonic decays of doubly charmed baryons  with $\Xi_c-\Xi_c'$ mixing}
\author{Chao-Qiang Geng, Xian-Nan Jin, Chia-Wei Liu, Xiao Yu and Ao-Wen Zhou\\}

%\email[]{Your e-mail address}
%\homepage[]{Your web page}
%\thanks{}
%\altaffiliation{}
\affiliation{School of Fundamental Physics and Mathematical Sciences, Hangzhou Institute for Advanced Study, UCAS, Hangzhou 310024, China\\
	University of Chinese Academy of Sciences, 100190 Beijing, China
	\vspace{0.6cm}} 

%Collaboration name if desired (requires use of superscriptaddress
%option in \documentclass). \noaffiliation is required (may also be
%used with the \author command).
%\collaboration can be followed by \email, \homepage, \thanks as well.
%\collaboration{}
%\noaffiliation

\date{\today}

\begin{abstract}
We study the $\Xi_c- \Xi_c'$ mixing effects in the semileptonic decays the doubly charm baryons of $\Xi_{cc}$. We focus on the ratio of ${\cal R}(\theta_c) \equiv {\cal B}( \Xi_{cc} \to \Xi_c'  e^+ \nu_e)/ {\cal B}( \Xi_{cc} \to \Xi_c e^+ \nu_e) $ and find that $({\cal R}(\theta_0),{\cal R}(- \theta_0)) =(0.46 \pm 0.01,7.33 \pm 0.23)$ with $\theta_0 = 0.137\pi$, which are in  sharp contrast to ${\cal R}(0)=2.15\pm0.11$ without the mixing. The ratio is enhanced~(suppressed) by a factor of four for a  negative~(positive) $\theta_c$. In addition,  the polarization asymmetries of $\Xi_c^{(\prime)}$ are found to be $\alpha(-\theta_0) = 0.32 ~(-0.76)$ and $\alpha(\theta_0) = -0.82~(-0.38)$. As ${\cal R}$  and $\alpha $ are  highly sensitive to $\theta_c$ and unaffected by the $W$-exchange contributions, they provide excellent opportunities to determine $\theta_c$  in the ongoing experiments.
\end{abstract}

\maketitle

The baryonic semileptonic decays provide excellent opportunities to examine the baryon wave functions, which are decomposed into four parts: color, spatial, spin and  flavor. The color part is trivial in three quark systems, while the spatial part is usually assumed to be symmetric in the low-lying baryons.  In the low-lying spin-half baryons, the spin and flavor are known to be entangled, and
together they are named as the spin-flavor configurations. They are independent of the quark model but nontrivial when the three quarks possess different flavors.  

Recently,  the mixing  of $\Xi_c$ and $\Xi_c'$, given by 
\begin{eqnarray}\label{angles}
&&\left(
\begin{array}{c}
     \Xi_c \\
     \Xi_c' 
\end{array}
\right) 
= \left(
\begin{array}{cc}
    \cos\theta_c  & \sin \theta_c  \\
    -\sin \theta_c &  \cos \theta_c 
\end{array}
\right)\left(
\begin{array}{c}
     \Xi_c^{\bm{\overline{3}}} \\
     \Xi_c^{\bm{6}}
\end{array}
\right) \,,~~~|\theta_c| = \theta_0 = 0.137(5) \pi\,,
\end{eqnarray}
has been proposed~\cite{Mixing}, where $\theta_c$ is determined  from  the mass relations~\cite{Jenkins:1996rr}
to the precision of $m_s^2/(m_cN^2_c)$ with  $m_{s,c}$ the quark masses  and $N_c$ the color number.
In Eq.~(\ref{angles}),  
$\Xi_c^{(\prime)}$ represent the physical baryons,  and $\overline{{\bf 3}}$ and ${\bf 6}$ are the representations under the $SU(3)$ flavor symmetry, corresponding to the light quark spins are in antisymmetric and symmetric, respectively.
With the mixing, we are able to explain the large $SU(3)_F$ violation in the singly charm baryon decays~\cite{Belle:2021crz, ALICE:2021bli}. Though the mixing effects are second order of $\theta_c$, the effects are still sizable. However, it forbids us to  fix the sign of $\theta_c$ from the singly charmed baryon semileptonic decays.
Therefore, the sign of $\theta_c$ remains as an open question. 

For the nonleptonic decays, attempts~\cite{Liu:2022igi, Ke:2022gxm} have been made in $\Xi_{cc}$, pounding on the ratio measured at LHCb~\cite{LHCb:2022rpd}
\begin{equation}
{\cal R}^{\text{non}} = \frac{{\cal B}(\Xi_{cc}^{++}\to \Xi_{c}^{\prime+} \pi^+)}{{\cal B}(\Xi_{cc}^{++} \to \Xi_{c}^{+} \pi^+)}= 1.42\pm0.17\pm0.10\,,
\end{equation}
where the first and second uncertainties are statistical and systematic, respectively. 
It disagrees with all predictions in the literature before the measurements~\cite{Wang:2017mqp,Cheng:2020wmk,Sharma:2017txj,Gerasimov:2019jwp,Gutsche:2018msz,Ke:2019lcf,Shi:2019hbf,Ivanov:2020xmw}, where ideal configurations were assumed.
The puzzle was partly resolved by the mixing, first founded in Ref.~\cite{Ke:2022gxm}. 
As both the denominator and numerator are affected by the mixing,  it  has a sizable impact on the ratio. 
However, the factorization ansatz has been assumed with $\theta_c$  treated as a free parameter in Ref.~\cite{Ke:2022gxm}. 
To overcome the weak point, two of the authors of this work~(Geng and Liu) have included the nonfactorizable contributions~\cite{Liu:2022igi}, and  found that the mixing angle in Eq.~(\ref{angles}) can explain ${\cal R}^{\text{non}}$ based on the methodology developed in Ref.~\cite{Cheng:2020wmk}.
The mixing effects modify  ${\cal R}^{\text{non}}$ largely from 6.74 to 1.45 with a negative angle of   $\theta_c$. 
We note that the same methodology has shown to be suitable in the heavy-flavor-conserving and $\Lambda_c^+$ nonleptonic decays~\cite{Cheng:2018hwl, Cheng:2022jbr}. 

Nevertheless, due to the hadron uncertainties, the sign of $\theta_c$ still requires a further confirmation, which is the main purpose of this work. 
To this end, we revisit our previous study in Ref.~\cite{Geng:2022uyy}, where $\theta_c = 0$ was assumed. The formalism we adopt here is  the same as Ref.~\cite{Geng:2022uyy} except  the spin-flavor configurations given in  Eq.~(\ref{angles}).  The modifications affect exclusively in the spin-flavor overlappings, defined as~\cite{Liu:2022pdk}
\begin{eqnarray}\label{const}
&&N^{(\prime)}_{\text{flip}} = \Big
\langle \Xi^{(\prime)}_c  ,J_z =\frac{1}{2} \Big | s^\dagger \sigma_z  c \Big| \Xi_{cc} ,J_z =\frac{1}{2} \Big \rangle \,,\nonumber\\&&
N_{\text{unflip}} ^{(\prime)} = 
\Big \langle \Xi^{(\prime)}_c  ,J_z =\frac{1}{2} \Big | s^\dagger   c \Big| \Xi _{cc} ,J_z =\frac{1}{2} \Big  \rangle\,,
\end{eqnarray}
where $ s^\dagger \sigma_z c = s^\dagger_\uparrow c_\uparrow - s^\dagger_\downarrow c_\downarrow $ and $s^\dagger c  = s^\dagger_\uparrow c_\uparrow +  s^\dagger_\downarrow c_\downarrow \,, $  while  
the baryon states in Eq.~(\ref{const}) are in the nonrelativistic constituent quark limit. 
We adopt the homogeneous bag model as it provides  simple spatial wave functions, leaving clarity to the spin-flavor configurations. Plugging Eq.~(\ref{angles}) into Eq.~(\ref{const}), we find  that 
\begin{eqnarray}
&&\left(
\begin{array}{c}
N_{\text{(un)flip}} \\
N_{\text{(un)flip}}'
\end{array}
\right) 
= \left(
\begin{array}{cc}
    \cos\theta_c  & \sin \theta_c  \\
    -\sin \theta_c &  \cos \theta_c 
\end{array}
\right)\left(
\begin{array}{c}
N_{\text{(un)flip}}^{\overline{\bf 3}}\\
N_{\text{(un)flip}}^{\bf 6}
\end{array}
\right),
\end{eqnarray}
resulting in that 
\begin{eqnarray}\label{eq5}
&&N_{\text{unflip}}^{\overline{{\bf 3}}, {\bf 6} } = \left(\frac{\sqrt{6}}{2}, \frac{\sqrt{6}}{6}\right)\,,~
N_{\text{flip}}^{\overline{{\bf 3}}, {\bf 6} } = \left(\frac{\sqrt{2}}{2}, \frac{5\sqrt{6}}{6}\right)\,.
\end{eqnarray}
 In the following,  we  omit the small uncertainty of  $\theta_0$ and simply take $\theta_0 = 0.137\pi$ in Eq.~(\ref{angles}).

In analogy to ${\cal R}^{non}$, 
we define the semileptonic version of the ratio
\begin{equation}\label{ratio}
    \mathcal{R} (\theta_c )= 
    \frac{\Gamma'}{\Gamma} = 
    \frac{\Gamma(\Xi_{cc} \to \Xi^{\prime }_{c}e^+ \nu_e)}{\Gamma(\Xi_{cc} \to \Xi_c e^+ \nu_e )}\,.
\end{equation}
Furthermore, the polarization asymmetries of $\Xi_c^{(\prime)}$ 
are defined as 
\begin{equation}\label{polarization asymmetry}
\alpha^{(\prime)} = \frac{
\Gamma^{(\prime)}( \lambda =1/2  ) - \Gamma^{(\prime)}( \lambda =-1/2  )
}{\Gamma^{(\prime)}( \lambda =1/2  ) + \Gamma^{(\prime)}( \lambda =-1/2  )}\,,
\end{equation}
where $\lambda$ are the helicities of $\Xi_c^{(\prime)}$.

\begin{figure}[b]
	\includegraphics[width=0.47\linewidth]{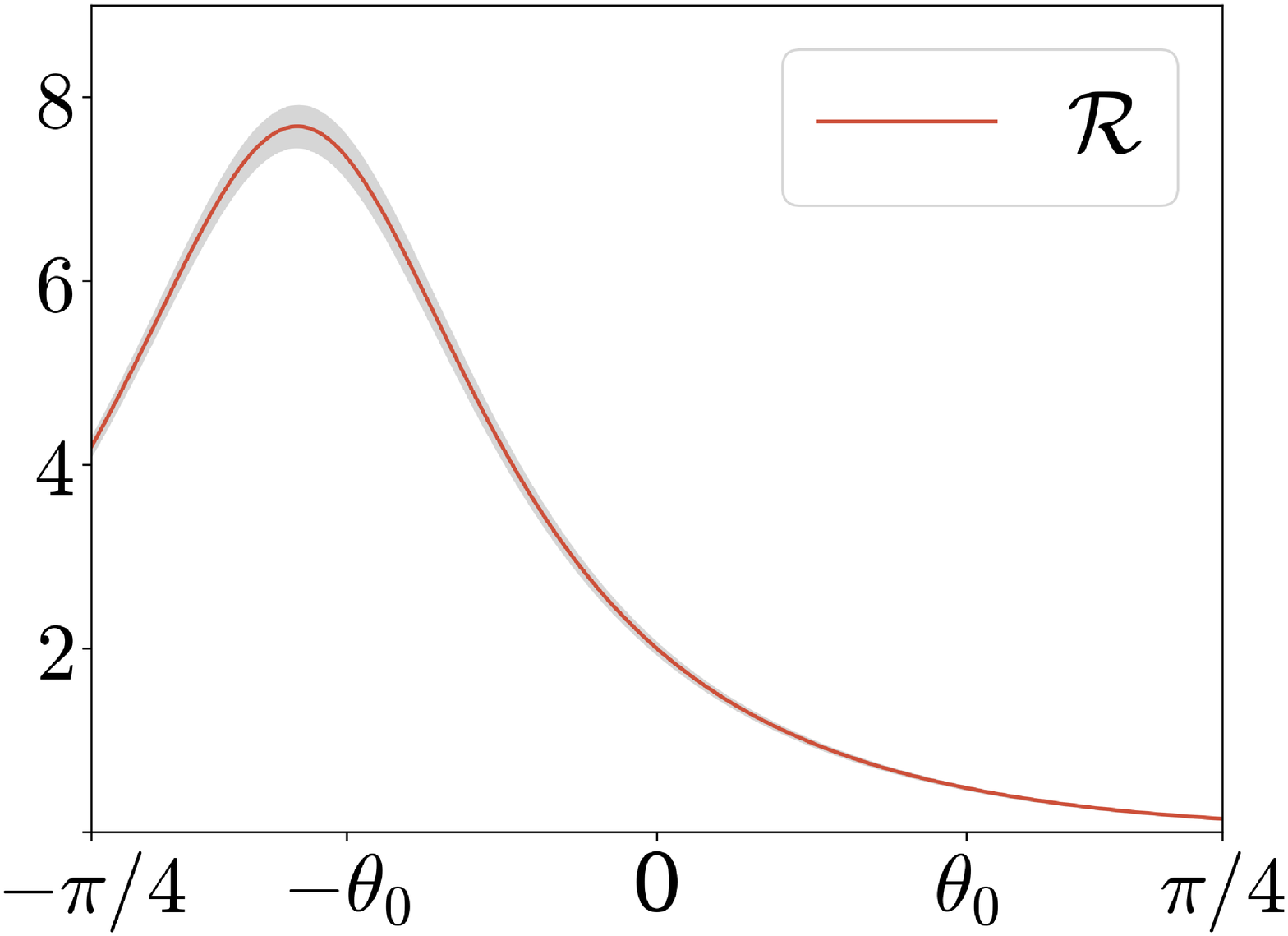}  	 	
	\includegraphics[width=0.51 \linewidth]{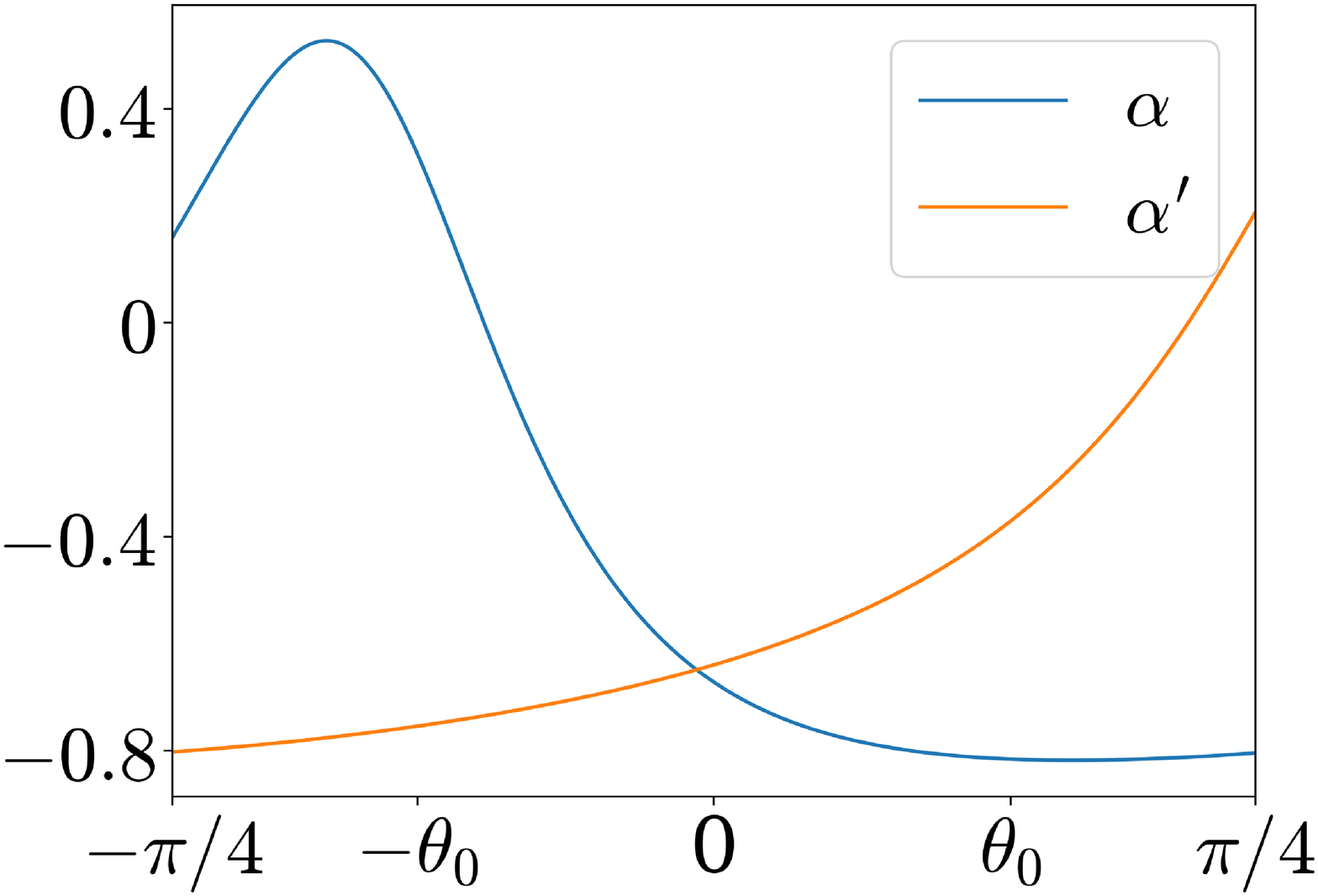}  	
	\caption{ 
		${\cal R}$ and $\alpha^{(\prime)}$ versus $\theta_c$.
 }
	\label{PA3} 
\end{figure}

The dependencies of  $\mathcal{R}$ and  $\alpha^{(\prime)}$ on $\theta_c$ are depicted in Fig.~\ref{PA3}, and 
their explicit values for  $\theta_c \in \{\pm\pi/4 ,\pm\theta_0,0\}$ are presented in TABLE~\ref{form2}. 
The uncertainties of $\alpha^{(\prime)}$ are  tiny and can be neglected in the considered precision. 
It shows that the physical quantities depend heavily on $\theta_c$. Explicitly,  ${\cal R}$ varies from $0.45$ to $7.56$ in the range of $\theta_c = \pm \theta_0$,  and  the sign of $\alpha$ is flipped for $\theta_c = -\theta_0$.  
The large deviations  can be   traced back to the interference between $N_{\text{(un)flip}}^{\overline{{\bf 3}}, {\bf 6}}$. 
In particular, a positive (negative) $\theta_c$ leads to  constructive (destructive)   and destructive (constructive) interference in  $N_{\text{(un)flip}}$ and  $N_{\text{(un)flip}}'$, respectively. 
 With this feature, we can identify the sign of $\theta_c$ once the experimental value of ${\cal R}$ is given. It is clear that  a negative $\alpha $ is a smoking gun of $\theta_c\neq 0$. 

\begin{table}[t]
\caption{$\Gamma^{(\prime)}$ and $\alpha^{(\prime)}$ in units of $10^{-5}$ eV and $\%$, respectively.}
\label{form2}
\begin{tabular}{c|ccccc}
% \begin{tabular}{c|ccccc}
\hline
\hline
$\theta_c$
&\multicolumn{1}{c}{$-\pi/4 $ }
&\multicolumn{1}{c}{$-\theta_0$}
&\multicolumn{1}{c}{$0$ }
&\multicolumn{1}{c}{$\theta_0$ }
&\multicolumn{1}{c}{$\pi/4  $ }
\\
\hline 
$\Gamma' $ &$ 12.7(13)$ &$13.5(14)$  &$ 10.9(8) $&$ 5.64(58) $  &$ 2.34(24) $ \\
$\Gamma$  &$ 3.04(38) $  &$ 1.85(25) $  &$ 5.11(64) $&$ 11.8(15) $&$ 16.2(21)$\\
${\cal R} $                             &$4.19(10)$    &$7.33(23)$    &$ 2.15(11) $&$0.46(1) $  &$0.14(0)$\\
\hline 
$\alpha$ & $15$&$32$ & $-68$&$-82$ &$-81$\\
$\alpha'$& $-80$&$-76$ &$-65$& $-38$&$-20$\\
\hline
\hline
\end{tabular}
\end{table}

We note that with $\theta_c=0$, ${\cal R}$ are found to be 
$2.39,$ $1.67,$ and $ 1.64$
from  the 
MIT bag model~\cite{Perez-Marcial:1989sch}, heavy quark spin symmetry~\cite{Albertus:2011xz}, and  light-front quark model~\cite{Hu:2020mxk},  respectively.  We emphasize that 
the future experimental value  of ${\cal R} >7$ $({\cal R}<0.5)$ will be a clear evidence of the  $\Xi_c-\Xi_c'$ mixing with negative (positive) $\theta_c$.

In sum, we have studied the $\Xi_c-\Xi_c'$ mixing effects in $\Xi_{cc} \to \Xi_c^{(\prime)} e^+ \nu_e$. We have found that ${\cal R} = 0.46\pm 0.01$~$(7.33\pm 0.23)$ for  $\theta_c=(-) \theta_0$, which is about four times smaller~(larger) than ${\cal R} = 2.15\pm 0.11$ without the mixing of  $\theta_c =0$.  Notably, with $\theta_c = -\theta_0$  the polarization asymmetry of $\Xi_c$  flips sign, given as $\alpha = 0.32$. 
We  recommend the future experiments to measure ${\cal R}$ and $\alpha^{(\prime)}$, since  $\alpha<0$ and  ${\cal R} > 7 $  are  smoking guns of the $\Xi_c - \Xi_c'$ mixing with $\theta_c <0$, whereas ${\cal R}<0.5$ indicates that  $\theta_c >0$. 
As the decays are highly sensitive to $\theta_c$ and  unaffected by the $W$-exchange contributions,  they  provide  the  most now known  reliable ways to probe $\theta_c$.

\begin{acknowledgments}
	This work is supported in part by the National Key Research and Development Program of China under Grant No. 2020YFC2201501 and  the National Natural Science Foundation of China (NSFC) under Grant No. 12147103.
\end{acknowledgments}


\begin{thebibliography}{99}

\bibitem{Mixing}
C.~Q.~Geng, X.~N.~Jin and C.~W.~Liu,
%``Resolving puzzle in $\Xi_c^0\to \Xi^-e^+\nu_e$ with $\Xi_c-\Xi_c'$ mixing,''
arXiv:2210.07211 [hep-ph].



\bibitem{Jenkins:1996rr}
E.~E.~Jenkins,
%``Heavy baryon masses in the 1/m(Q) and 1/N(c) expansions,''
Phys. Rev. D \textbf{54}, 4515 (1996);
E.~E.~Jenkins,
%``Update of heavy baryon mass predictions,''
Phys. Rev. D \textbf{55}, 10 (1997);
E.~E.~Jenkins,
%``Model-Independent Bottom Baryon Mass Predictions in the 1/N($c$) Expansion,''
Phys. Rev. D \textbf{77}, 034012 (2008).



\bibitem{Belle:2021crz}
Y.~B.~Li \textit{et al.} [Belle],
Phys. Rev. Lett. \textbf{127}, 121803 (2021).
%1

\bibitem{ALICE:2021bli}
S.~Acharya \textit{et al.} [ALICE],
Phys. Rev. Lett. \textbf{127}, 272001 (2021).
%2

\bibitem{Ke:2022gxm}
H.~W.~Ke and X.~Q.~Li,
Phys. Rev. D \textbf{105}, 096011 (2022).
%3

\bibitem{Liu:2022igi}
C.~W.~Liu and C.~Q.~Geng,
%``Nonleptonic decays of $\Xi_{cc}\to \Xi_c \pi$ with $\Xi_c-\Xi_c'$ mixing,''
arXiv:2211.12960 [hep-ph].
%4

\bibitem{LHCb:2022rpd}
R.~Aaij \textit{et al.} [LHCb],
JHEP \textbf{05}, 038 (2022).
%5
\bibitem{Wang:2017mqp}
W.~Wang, F.~S.~Yu and Z.~X.~Zhao,
Eur. Phys. J. C \textbf{77}, 781 (2017).
%6

\bibitem{Sharma:2017txj}
N.~Sharma and R.~Dhir,
Phys. Rev. D \textbf{96}, 113006 (2017).
%7

\bibitem{Gerasimov:2019jwp}
A.~S.~Gerasimov and A.~V.~Luchinsky,
Phys. Rev. D \textbf{100}, 073015 (2019).
%8

\bibitem{Gutsche:2018msz}
T.~Gutsche, M.~A.~Ivanov, J.~G.~K\"orner, V.~E.~Lyubovitskij and Z.~Tyulemissov,
Phys. Rev. D \textbf{99}, 056013 (2019).
%9

\bibitem{Cheng:2020wmk}
H.~Y.~Cheng, G.~Meng, F.~Xu and J.~Zou,
Phys. Rev. D \textbf{101}, 034034 (2020).
%10

\bibitem{Shi:2019hbf}
Y.~J.~Shi, W.~Wang and Z.~X.~Zhao,
Eur. Phys. J. C \textbf{80}, 568 (2020).
%11

\bibitem{Ivanov:2020xmw}
M.~A.~Ivanov, J.~G.~K\"orner and V.~E.~Lyubovitskij,
Phys. Part. Nucl. \textbf{51}, 678-685 (2020).
%12

\bibitem{Ke:2019lcf}
H.~W.~Ke, F.~Lu, X.~H.~Liu and X.~Q.~Li,
Eur. Phys. J. C \textbf{80}, 140 (2020).
%13


\bibitem{Cheng:2018hwl}
H.~Y.~Cheng, X.~W.~Kang and F.~Xu,
%``Singly Cabibbo-suppressed hadronic decays of $\Lambda_c^+$,''
Phys. Rev. D \textbf{97},  074028 (2018).

\bibitem{Cheng:2022jbr}
H.~Y.~Cheng, C.~W.~Liu and F.~Xu,
%``Heavy-flavor-conserving hadronic weak decays of charmed and bottom baryons: An update,''
Phys. Rev. D \textbf{106},  093005 (2022).



\bibitem{Geng:2022uyy}
C.~Q.~Geng, C.~W.~Liu, A.~Zhou and X.~Yu,
arXiv:2211.04372 [hep-ph].
%14

\bibitem{Liu:2022pdk}
C.~W.~Liu and C.~Q.~Geng,
%``Center of mass motion in bag model,''
arXiv:2205.08158 [hep-ph].
%15

\bibitem{Albertus:2011xz}
C.~Albertus, E.~Hern\'andez and J.~Nieves,
Phys. Lett. B \textbf{704}, (2011).
%16

\bibitem{Perez-Marcial:1989sch}
R.~Perez-Marcial, R.~Huerta, A.~Garcia and M.~Avila-Aoki,
Phys. Rev. D \textbf{40}, 2955 (1989)
[erratum: Phys. Rev. D \textbf{44}, 2203 (1991)].

\bibitem{Albertus:2011xz}
C.~Albertus, E.~Hern\'andez and J.~Nieves,
Phys. Lett. B \textbf{704}, 499 (2011).


\bibitem{Hu:2020mxk}
X.~H.~Hu, R.~H.~Li and Z.~P.~Xing,
Eur. Phys. J. C \textbf{80}, 320 (2020).
%17




\end{thebibliography}
\end{document}